\documentclass[11pt]{article}
\usepackage{rpmacros}
\usepackage[dvipsnames,svgnames,x11names,hyperref]{xcolor}
\usepackage{amsthm}
\usepackage{xfrac}
\usepackage{mathpazo}
\usepackage{gitinfo}
\usepackage{enumitem}
\usepackage{fullpage}
\usepackage{thmtools}
\usepackage[T1]{fontenc} 
\usepackage[colorlinks]{hyperref}
\hypersetup{
  linkcolor=[rgb]{0,0,0.4},
  citecolor=[rgb]{0, 0.4, 0},
  urlcolor=[rgb]{0.6, 0, 0}
}
\usepackage{lipsum}
\usepackage{mathrsfs,bbm}
\usepackage{bm}
\usepackage{todonotes}
\usepackage{tikz}
\usetikzlibrary{decorations.pathreplacing,backgrounds}
\usepackage{multirow}
\usepackage{setspace}

\usepackage[linesnumbered,vlined,ruled]{algorithm2e}

\usepackage{amsthm}
\usepackage{thmtools,thm-restate}
\usepackage[nameinlink,capitalise]{cleveref}

\numberwithin{equation}{section}
\declaretheoremstyle[bodyfont=\it,qed=\qedsymbol]{noproofstyle}

\declaretheorem[name=Observation,numberlike=equation]{observation}

\declaretheorem[name=Observation,numbered=no]{observation*}

\declaretheorem[numberlike=equation]{theorem}

\declaretheorem[name=Theorem,numbered=no]{theorem*}

\declaretheorem[numberlike=equation]{lemma}
\declaretheorem[name=Lemma,numbered=no]{lemma*}
\declaretheorem[numberlike=equation,style=noproofstyle,name=Lemma]{lemmawp}

\declaretheorem[name=Corollary,numbered=no]{corollary*}

\declaretheorem[name=Proposition,numbered=no]{proposition*}

\declaretheorem[name=Claim,numbered=no]{claim*}

\declaretheorem[name=Conjecture,numbered=no]{conjecture*}

\declaretheorem[name=Question,numbered=no]{question*}

\declaretheoremstyle[bodyfont=\it,qed=$\lozenge$]{defstyle} 

\declaretheorem[numberlike=equation,style=defstyle]{definition}
\declaretheorem[unnumbered,name=Definition,style=defstyle]{definition*}

\declaretheorem[unnumbered,name=Example,style=defstyle]{example*}

\declaretheorem[unnumbered,name=Notation=defstyle]{notation*}

\declaretheorem[unnumbered,name=Construction,style=defstyle]{construction*}

\declaretheorem[unnumbered,name=Remark,style=defstyle]{remark*}

\renewcommand{\phi}{\varphi}
\renewcommand{\epsilon}{\varepsilon}

\newcommand{\SP}{\Sigma\Pi}
\newcommand{\SPsize}[1]{(\SP)^k\operatorname{-size}}

\usepackage{nth}
\usepackage{intcalc}
\usepackage{etoolbox}
\usepackage{xstring}

\usepackage{ifpdf}
\ifpdf
\else
\usepackage[quadpoints=false]{hypdvips}
\fi

\newcommand{\ECCCurl}[2]{http://eccc.hpi-web.de/report/\ifnumcomp{#1}{>}{93}{19}{20}#1/#2/}

\newcommand{\shortECCC}[2]{\texttt{\href{http://eccc.hpi-web.de/report/\ifnumcomp{#1}{>}{93}{19}{20}#1/#2/}{eccc:TR#1-#2}}}

\newcommand{\parseECCC}[1]{
\StrSubstitute{#1}{TR}{}[\tmpstring]%
\IfSubStr{\tmpstring}{/}{ 
\StrBefore{\tmpstring}{/}[\ecccyear]%
\StrBehind{\tmpstring}{/}[\ecccreport]%
}{
\StrBefore{\tmpstring}{-}[\ecccyear]%
\StrBehind{\tmpstring}{-}[\ecccreport]%
}%
\shortECCC{\ecccyear}{\ecccreport}}

\newcommand{\homog}{\operatorname{Hom}}

\newcommand*\samethanks[1][\value{footnote}]{\footnotemark[#1]}

\newif\ifblind
\blindfalse

\newif\ifdraft
\draftfalse
\ifdraft
\newcommand{\RPnote}[1]{\textcolor{BrickRed}{\guillemotleft RP: #1 \guillemotright}}
\newcommand{\MKnote}[1]{\textcolor{Orange}{\guillemotleft MK: #1 \guillemotright}}
\newcommand{\VRnote}[1]{\textcolor{Blue}{\guillemotleft VR: #1 \guillemotright}}

\newcommand{\gitinfonotecolour}{Gray}
\newcommand{\easteregg}{}
\else
\newcommand{\RPnote}[1]{}
\newcommand{\MKnote}[1]{}
\newcommand{\VRnote}[1]{}
\newcommand{\gitinfonotecolour}{white}
\newcommand{\easteregg}{}
\fi

\newcommand{\ignore}[1]{}
\newcommand{\gitinfonote}{git info:~\gitAbbrevHash\;,\;(\gitAuthorIsoDate)\; \;\gitVtag}

\newcommand{\Res}[3]{\ensuremath{\operatorname{Res}_{#1}(#2,#3)}}

\renewcommand{\vec}[1]{\ensuremath{\bm{#1}}}

\newcommand{\Ring}{\ensuremath{\mathcal{R}}}

\newcommand{\C}{\mathbb{C}}
\newcommand{\num}{\ensuremath{\mathrm{num}}}
\newcommand{\den}{\ensuremath{\mathrm{den}}}

\title{Towards Deterministic Algorithms for Constant-Depth Factors of Constant-Depth Circuits}

\ifblind
\else
\author{
    {Mrinal Kumar \thanks{Tata Institute of Fundamental Research, Mumbai, India. Email: \texttt{\{mrinal, varun.ramanathan, ramprasad\}@tifr.res.in}.  Research supported by the Department of Atomic Energy, Government of India, under project 12-R{\&}D-TFR-5.01-0500.}}
    \and
    {Varun Ramanathan{\samethanks[1]}}
    \and
    {Ramprasad Saptharishi{\samethanks[1]}}
    \and
    {Ben Lee Volk\thanks{Efi Arazi School of Computer Science, Reichman University, Israel. Email: \texttt{benleevolk@gmail.com}. The research leading to these results has received funding from the Israel Science Foundation (grant number 843/23).}}
}
\fi
\date{}

\begin{document}

\maketitle

\begin{abstract}
We design a deterministic subexponential time algorithm that takes as input a multivariate polynomial $f$  computed by a constant-depth circuit over rational numbers, and outputs a list $L$ of circuits (of unbounded depth and possibly with division gates) that contains all irreducible factors of $f$ computable by constant-depth circuits. This list $L$ might also include circuits that are spurious: they either do not correspond to factors of $f$ or are not even well-defined, e.g. the input to a division gate is a sub-circuit that computes the identically zero polynomial. 

The key technical ingredient of our algorithm is a notion of the \emph{pseudo-resultant} of $f$ and a factor $g$, which serves as a proxy for the resultant of $g$ and $f/g$, with the advantage that the circuit complexity of the pseudo-resultant is comparable to that of the circuit complexity of $f$ and $g$. 
This notion, which might be of independent interest, together with the recent results of Limaye, Srinivasan and Tavenas \cite{LST21} helps us derandomize one key step of multivariate polynomial factorization algorithms --- that of deterministically finding a good starting point for Newton Iteration for the case when the input polynomial as well as the irreducible factor of interest have small constant-depth circuits. 
\end{abstract}

\section{Introduction}
\label{sec:intro}

Algorithms for polynomial factorization is an area in which the field of computer algebra has been remarkably successful. Unlike the analogous and notoriously hard problem of \emph{integer} factorization, a sequence of works in the last few decades provided clever and efficient algorithms for factorizing \emph{polynomials}. These algorithms work in many different settings: finite and infinite fields, univariate and multivariate, white- vs.\ black-box, and so on \cite{Forbes-Shpilka-survey,GG13}.

One of the first questions in the design of factorization algorithms is how is the input represented (and, for that matter, what are the requirements regarding the representation of the output). An $n$-variate polynomial of degree $d$ has potentially $\binom{n+d}{d}$ monomials, so representing it as a list of coefficients under some ordering of the monomials (the so-called \emph{dense} representation) would cause the input to be of roughly that size. Even in this model, where the running time is allowed to be polynomial in $\binom{n+d}{d}$, factorization is not a trivial task. Some polynomials, however, have more efficient representations, the most natural of which being an \emph{algebraic circuit}. Such a circuit is a directed acyclic graph, whose inputs are labeled by variables $x_1, \ldots, x_n$ and field elements, and internal nodes labeled by the arithmetic operations $+,-,\times,\div$. Such a circuit naturally represents (or computes) the polynomial(s) computed in its output gate(s).

A series of influential works culminated in efficient randomized algorithms by Kaltofen \cite{K89} and Kaltofen and Trager \cite{KT88} for factorization of polynomials given by algebraic circuits. These algorithms have found numerous applications in various fields (e.g., \cite{KI04, DSY09,S97, GS99}, to name a few, and see \cite{Forbes-Shpilka-survey} for more background) They also established the highly non-obvious, and perhaps surprising fact, that factors of polynomials computed by small circuits themselves have small circuits.\footnote{A more accurate statement would be that if $f$ is an $n$-variate polynomials of degree $\poly(n)$ that has a circuit of size $\poly(n)$, then any factor $g$ of $f$ has a circuit of size $\poly(n)$. In general, however, $n$-variate polynomials with circuits of size $\poly(n)$ can have exponential degree. For such polynomials, the questions of whether their factors are efficiently computable is still open and known as the \emph{factor conjecture} \cite[Conjecture 8.3]{Burgisser00}}. Faced with this fact, it is natural to wonder how far this connection extends: that is, suppose $\mathcal{C}$ is a class of algebraic circuits. Can we argue that factors of polynomials that have circuits in $\mathcal{C}$ themselves have circuits in $\mathcal{C}$? Such connections are known when $\mathcal{C}$ is a class that is powerful enough to support the arguments used in Kaltofen's proof, such as the class of polynomial size algebraic circuits or polynomial size algebraic branching programs \cite{ST21}. But much less is known when $\mathcal{C}$ is a weaker class, even for classes that we understand pretty well. For example, despite considerable effort \cite{BSV20} such a statement isn't known to hold when $\mathcal{C}$ is the class of depth-$2$ circuits of polynomial size (that is, circuits that compute sparse polynomials).

Further, algorithms of Kaltofen \cite{K89} and Kaltofen and Trager \cite{KT88} are randomized, and from a theoretical viewpoint this raises the intriguing problem of derandomizing them. However, as observed by Shpilka and Volkovich \cite{SV10}, obtaining a deterministic algorithm for factorization implies an efficient algorithm for the famous Polynomial Identity Testing (PIT) problem, which in turn implies circuit lower bounds \cite{KI04}. Indeed, $f(x_1, \ldots, x_n)$ is non-zero if and only if the polynomial $g(x_1, \ldots, x_n, y,z) = f(x_1, \ldots, x_n)+yz$ is irreducible, and given a circuit for $f$ one can easily obtain a circuit for $g$ of nearly the same size, and further, for almost any non-trivial class $\mathcal{C}$, a $\mathcal{C}$-circuit for $f$ implies a $\mathcal{C}$-circuit for $g$. Thus, we see that factorization of circuits from restricted classes $\mathcal{C}$ is at least as hard as polynomial identity testing for circuits from the same class.

This motivates studying the factorization problem for restricted classes $\mathcal{C}$ for which we can obtain non-trivial PIT algorithms. One can be even further encouraged by the fact that Kopparty, Saraf and Shpilka \cite{KSS14} proved an equivalence between derandomization of PIT and factorization: namely, they proved that a polynomial time deterministic algorithm for PIT would imply a polynomial time algorithm for factorization, by showing how to use PIT to derandomize Kaltofen's algorithm. However, as before (and for the same reason), their argument breaks down when specialized to circuits from a restricted class $\mathcal{C}$: even when trying to factorize polynomials from a weak class $\mathcal{C}$, Kaltofen's algorithm (and consequentially, its conditional derandomization by Kopparty, Saraf and Shpilka \cite{KSS15}) is forced to solve PIT instances for strong classes, such as the class of polynomial size algebraic branching program. This leaves open the interesting question of deterministically factoring polynomials computed by limited classes of circuits.

In fact, even for simpler questions such as irreducibility testing of polynomials, or checking whether one polynomial divides another, it is hard to obtain deterministic algorithms, even under restricting assumptions on the complexity of the polynomials involved. A step in this direction was taken by Forbes \cite{Forbes15}, who showed how to deterministically test whether a quadratic polynomial divides a sparse polynomial.

More recently, in \cite{KRS23} the first three authors of this paper obtained a deterministic quasipolynomial time algorithm that outputs all constant-degree factors of a sparse polynomial. Even more generally, they showed how to use recent PIT results for constant-depth circuits \cite{LST21} in order to compute all constant-degree factors of constant-depth circuits, albeit in subexponential time. 

Polynomials of constant degree are in particular sparse and hence as a natural possible generalization of this result, one would like to be to deterministically output all sparse factors of constant-depth circuits, and even more generally, all constant-depth factors of constant-depth circuits. In this work, we make a step in this direction.

\subsection{Our Results}
\label{sec:intro:results}

Our main result in this work is the following theorem.

\begin{theorem}[Subexponential list containing constant-depth factors of constant-depth circuits]\label{thm:intro:main}
  Let $\Q$ be the field of rational numbers and $\epsilon > 0$, $\Delta_1, \Delta_2 \in \N$ be arbitrary constants. 
  
  Then, there is a deterministic algorithm that takes as input an algebraic circuit $C \in (\SP)^{(\Delta_1)}$ of size $s$, bit-complexity $t$ and degree $D$, along with a size parameter $m$, and outputs a list of circuits $L = \inbrace{C_1, \dots, C_r}$ with the following properties.
   \begin{enumerate}
    \item Each $C_i \in L$ is an arithmetic circuit of size $\poly(s,m,D)$ and bit-complexity $\poly(t,s,D,m)$, with division gates.
    \item For every irreducible factor $g$ of $C$ with a size $m$ $(\SP)^{(\Delta_2)}$-circuit computing it, there exists a $C_i \in L$ such that $C_i$ computes $g$.
   \end{enumerate} 
   The size of the list $L$ as well as the running time of the algorithm are bounded above by $O(smDt)^{O((smDt)^{\epsilon})}$. 
\end{theorem}

We remark that indeed \cref{thm:intro:main} doesn't completely solve the problem of finding sparse (or more generally, low depth) factors of low depth circuits. It merely outputs a ``short'' (subexponential) list $L$ of candidate polynomials such that every irreducible factor of the input is contained in $L$. As explained in \cref{sec:overview}, for technical reasons, we currently do not know how to deterministically prune the list $L$ to obtain the true factors.

Yet another technicality is that our algorithm needs a size bound on the size of a circuit for the factors we care about because no structural guarantees are known for the factors of constant-depth circuits. For instance, we do not know if the factors of a constant-depth circuit also have small constant-depth circuits.

\paragraph*{A slightly more general statement.}

Our techniques in the proof of \cref{thm:intro:main} give a slightly more general statement connecting the questions of deterministic polynomial factorization and deterministic polynomial identity testing. Let $\mathcal C$ be a class of circuits that are somewhat rich in the sense that if a circuit $C$ is in $\mathcal C$, then the circuit $C'$ obtained from $C$ by a change of basis continues to be in $\mathcal C$, and for circuits $C_1, C_2, \ldots, C_n$ in $\mathcal C$, and a circuit $B(y_1, y_2, \ldots, y_n)$ of constant depth, the circuit $B(C_1, C_2, \ldots, C_n)$ continues to be in $\mathcal C$. For instance, the class of polynomials computable by small constant-depth circuits, or by small formulas satisfy the above property. Our techniques in the proof of \cref{thm:intro:main} generalize to such classes and give a deterministic reduction from the question of computing irreducible factors of a polynomial $f \in \mathcal{C}$ that are also computable in $\mathcal C$ to the question of polynomial identity testing of polynomials (of slightly larger complexity) in $\mathcal C$; the major caveat being that the algorithm will only output a list containing all such irreducible factors of interest, but might also contain some spurious circuits. Moreover, these circuits in the output list are potentially of unbounded depth, can contain division gates and some of these spurious circuits in the list might not even be well defined, i.e., they involve division by an identically zero polynomial. In contrast to this, the prior known connections between PIT and polynomial factorization, most notably  in the work of Kopparty, Saraf and Shpilka \cite{KSS15} gives a truly deterministic polynomial factorization algorithm, given access to an oracle for PIT. However, even if the input to the polynomial factorization algorithm is very structured (for instance, is a sparse polynomial), the PIT instances generated in the process appear to be very general. 

In spite of this slight generality, for a clean and complete presentation of the results, we only work with constant-depth circuits in the rest of the paper.

\paragraph*{Field dependence of our result.} For our results, we work over the field of rational numbers due to various technical reasons. Our proofs rely on derivative-based techniques like working with the Taylor expansion of a polynomial, and working over fields of characteristic zero ensures that derivatives do not inadvertently vanish. We also need a deterministic algorithm for univariate factorization over the underlying field as a subroutine for our algorithms. Moreover, we also need a non-trivial deterministic algorithm for polynomial identity testing of constant-depth circuits over the underlying field \cite{LST21}. The field of rational numbers satisfies all these requirements. 

\subsection{Open Problems}
\label{sec:intro:open}
We conclude this section with some open problems. 
\begin{itemize}
\item Perhaps the most natural question here would be to prune the list output in \cref{thm:intro:main} so that it has no circuits that do not correspond to true factors. One possible approach to do this would be to understand the complexity of the lifting process in polynomial factorization algorithms better, and show improved upper bounds on the complexity of the roots and factors and then do deterministic divisibility testing. At the moment, it is unclear to us if this approach is feasible. 

\item One special case of \cref{thm:intro:main} that is already very interesting is that of sparse polynomials. It would be interesting to see if \cref{thm:intro:main} can be improved for this case in any way. For instance, can the time complexity of the algorithm be reduced to quasipolynomial time, or can spurious factors in the output list be eliminated deterministically when the input is sparse?

\item In general, the question of what is the complexity of factors of simple polynomials (e.g. sparse polynomials, constant-depth circuits) is of great interest, yet poorly understood. We do not know whether such classes are closed under taking factors, and evidence in any direction, either towards or against such closure results, would be very interesting. 
\end{itemize}

\subsection*{Organization}
The rest of the paper is organized as follows. We discuss a high level overview of the proof in \cref{sec:overview} and introduce and discuss the notion of pseudo-resultant in \cref{sec:pseudo-resultant}. In \cref{sec:generating-list}, we build upon this notion to describe and present our final algorithms. 

As with many of the papers on polynomial factorization, we have a somewhat elaborate preliminaries section with many of the standard facts and results. To keep this paper self-contained, we include this section in the paper; however, to avoid obstructing the flow of the paper, we have moved most of the section to \cref{sec:prelims} and \cref{sec:building-blocks}. 

\section{Proof Overview}
\label{sec:overview}
In this section, we give a brief overview of the main ideas in our proof. But first, we set up some notation: $f$ denotes the input polynomial of degree $d$ that is given via a small algebraic circuit of size $s$ and depth $\Delta$ computing it. Furthermore, we assume that $f$ factors as $f = g^m\cdot h$, where $g$ is an irreducible polynomial and can be computed by a small constant-depth circuit, and $g$ and $h$ do not have a non-trivial GCD. Throughout the underlying field is assumed to be the field $\Q$ of rational numbers. We note that a priori, it is not clear if $h$ can also be computed by a small constant-depth circuit. However, this follows immediately from an observation in a recent prior work of Kumar, Saptharishi and Ramanathan \cite{KRS23}. 

 We now outline the sketch of our ideas. Our algorithm follows the general outline of a typical factorization algorithms in the literature (e.g., \cite{K89}). However, in their original formulations, some of these steps rely on randomness, and we elaborate how we get around this and implement the step in a deterministic way.   \\

\noindent
\textbf{Making $f$ monic: } As the first step of the algorithm, we make the polynomial $f$ monic in one of the variables, which we denote by $y$ here by applying an invertible linear transformation to the variables. This is typically done by choosing field constants $\veca = (a_1,a_2, \ldots, a_n) \in \F^n$ at random, and replacing the variable $x_i$ in the circuit for $f$ by the linear form $x_i + a_i y$. Since the underlying field is large enough, via the Schwartz-Zippel lemma, we get that the coefficient of  $y^d$ in $f(\vecx + \veca \cdot y)$ is a non-zero field constant with high probability, and up to a scaling by this field constant, we obtain a polynomial that is monic in $y$. For the ease of notation, we continue to denote this new polynomial by $f$. To derandomize this step, we note that the coefficient of $y^d$ in $f(\vecx + \veca \cdot y)$  is just the evaluation of the homogeneous component of $f(\vecx)$ of degree $d$ on the input $\veca$. But since $f$ has a size $s$, depth $\Delta$ circuit,  its degree $d$ homogeneous component has a circuit of size $s' = \poly(s,d)$ and depth $\Delta' = \Delta + O(1)$ (see \cref{lem:computing-hom-components}). Thus, if we try all possible inputs $\veca$ from a  hitting set for size $s'$, depth $\Delta'$ circuits, one of the vectors will have the desired property. A deterministic subexponential time construction of such a hitting set for constant-depth circuits was shown by Limaye, Srinivasan and Tavenas \cite{LST21}. 

For the rest of the algorithm, we view $f$ as a polynomial in $\Q[\vecx][y]$. We also note that since $f$ is monic in $y$, without loss of generality, $g$ and $h$ can be assumed to be monic in $y$ (\cref{lem: gauss lemma}).\\ 

\noindent
\textbf{Reducing the multiplicity of $g$ in $f$: } The remaining steps of the algorithm assume that the multiplicity of $g$ in $f$, denoted by the parameter $m$ here, is one. 
This can be assumed without loss of generality. Indeed, to reduce to this case, we run our algorithm on  all the partial derivatives of $f$ with respect to $y$, i.e. the polynomials $\{\frac{\partial^i f}{\partial y^i} : i \in \{0, 1, \ldots, d-1\}\}$; the observation being that $g$ is an irreducible factor of multiplicity one of at least one of these polynomials. Moreover, all these polynomials can be computed by a circuit of size $\poly(s,d)$ and depth $\Delta$ (\cref{lem: partial derivatives with respect to one variable}). So, it suffices to focus on recovering factors of multiplicity one of $f$. This observation was also used in \cite{KRS23}.\\

\noindent
\textbf{Finding a good starting point for Newton Iteration: }The goal now is to view $f$ as a univariate in $y$ with coefficients from $\Q[\vecx]$ and obtain a root $\Phi(\vecx)$ of $f$ in the ring $\Q[[\vecx]]$ of power series in $\vecx$. Moreover, in order to recover the factor $g$ of $f$ using this root, it must be the case that this is a root of $g$ (and hence a root of $f$). In fact, by standard techniques, it suffices to  recover an approximation of $\Phi$ with high enough accuracy, i.e., to recover  $\Phi_k(\vecx) := \Phi(\vecx) \mod \langle \vecx \rangle^k$ for some $k > 2d^2$. This is done via Newton iteration, where we start from the constant term (i.e. $\Phi_0 := \Phi(\vecx) \mod \langle \vecx \rangle$) of such a $\Phi$, and \emph{lift} it to obtain $\Phi_k$. While the lifting process is itself deterministic (see \cref{lem:newton-iteration-approx-root}), a crucial issue  with respect to the starting point of the lifting process is the following:  $\Phi_0$ must be a root of $f(\vecx,y) \bmod \langle \vecx \rangle$ of multiplicity one, and in fact must come from $g \bmod \langle \vecx \rangle$. This is typically done by translating the $\vecx$ variables to $\vecx + \veca$ such that the univariate polynomials $g(\veca, y)$ and $h(\veca, y)$ do not share a non-trivial GCD and $g(\veca, y)$ is square-free. This ensures that every root of $f(\veca, y) = g(\veca, y)h(\veca, y)$ that comes from $g(\veca, y)$ is a root of multiplicity one of $f(\veca, y)$ and hence is a good starting point for lifting via Newton Iteration. 

In typical factorization algorithms, such a point $\veca$ is found by first preprocessing the input $f$ such that it is square-free, and then setting $\veca$ to be a non-zero of the discriminant of $f$. Since the discriminant is an efficiently computable low-degree polynomial (assuming $f$ is), by the Schwartz-Zippel lemma $\veca$ can be chosen at random. This guarantees that $f(\veca, y) = g(\veca, y)h(\veca, y)$ is square-free and hence $g(\veca,y)$ and $h(\veca, y)$ don't have a non-trivial GCD. In our setting, we do not have the guarantee that $f(\vecx, y)$ is square-free since $h$ might not be square-free, but we know that $g$ is a multiplicity one irreducible factor of $f$. Moreover, we would like to find such a point $\veca$ deterministically. A serious quantitative issue with respect to derandomizing the application of the Schwartz-Zippel lemma here is that even though $f$ has a small constant-depth circuit, its discriminant is the determinant of a matrix of dimension roughly $2d$ with polynomial entries, a powerful model for which no non-trivial deterministic PIT algorithms are known. This is also the issue if we try to look at the resultant of $g$ and $h$ instead.

 One intriguing possibility  here would be to show that the discriminant or the resultant of polynomials with small constant-depth circuits have relatively small constant-depth circuits. However, we do not know how to show this. Instead, we come up with an alternative polynomial $R_{f,g}$, referred to as the \emph{pseudo-resultant} of $f$ and $g$,
and rely on its properties to arrive at our good starting point for the lifting process. In particular, we note that the complexity of $R_{f,g}$ is close to the complexity of $f$ and $g$. We then show that when $g$ is an irreducible factor of $f$ with multiplicity one, $R_{f,g}(\vecx,y) \not\equiv 0$. Moreover, using ideas from a divisibility testing algorithm of Forbes \cite{Forbes15}, we can show that for any $\veca \in \Q^n$ such that $R_{f,g}(\veca,y) \not\equiv 0$, there is a root of multiplicity one of $f(\veca, y)$ that is a root of $g(\veca,y)$ but not $h(\veca,y)$. This root becomes a good starting point for the lifting process. Furthermore, since $R_{f,g}$ is computable by a small constant-depth circuit, such a point $\veca$ can be obtained deterministically in subexponential time using the results in \cite{LST21}.

Intuitively, while the randomized algorithm for finding a good starting point for lifting uses the resultant to find an $\veca$ such that $g(\veca, y)$ and $h(\veca, y)$ do not have a non-trivial GCD, we observe that this is a bit of an overkill, and it suffices to find an $\veca$ such that the GCD of $g(\veca,y)$ and $h(\veca,y)$ is not equal to $g(\veca, y)$. Equivalently, it suffices to find an $\veca$ such that $g(\veca, y)$ does not divide $h(\veca, y)$. This happens to be a computationally easier condition to achieve deterministically via the pseudo-resultant. This is essentially the main technical idea in this work, and is formally discussed in \cref{sec:pseudo-resultant}.

One undesirable quantitative issue here is that the circuits obtained for $\Phi$ via Newton iteration, while being of polynomially bounded size, can have unbounded depth. This is one of the reasons that the output list of our algorithm can have spurious circuits not corresponding to factors of $f$.\\

\noindent
\textbf{Solving a linear system to obtain $g$: } Once we have obtained an approximate root $\Phi_k(\vecx)$ of $f(\vecx, y)$, we would like to recover the real factor $g(\vecx, y)$. If the $y$-degree of $g$ is known to be $d'$, then this is done via thinking of $g$ as $g(\vecx, y) = y^{d'} + \sum_{i = 0}^{d'-1}g_i(\vecx) y^i$, where $g_i$ is a formal variable, and imposing the constraint that 
\[
g(\vecx, \Phi_k(\vecx)) = \phi_k(\vecx)^{d'} + \sum_{i = 0}^{d'-1} g_i(\vecx) \Phi_k(\vecx)^i \equiv 0 \mod \langle \vecx \rangle^k \, .
\]
Furthermore, each $g_i$ can be written as $g_i(\vecx) = \sum_{j = 0}^d g_{i,j}(\vecx)$, where $g_{i,j}$ is the homogeneous component of $g_i$ of degree $j$. The constraint above translates naturally into a linear system in the variables $g_{i,j}$, and the factor $g$ does satisfy this system of equations. Furthermore, it can be shown that if $d'$ equals the $y$-degree of $g$ and $k > 2d^2$, then $g$ is the unique solution to this system of linear equations. Moreover, if we set a similar system with the $y$-degree of the factor of interest being set to something less than $d'$, then the resulting linear system has no solution. We rely on these properties of the linear system to obtain an algebraic circuit for $g$. Essentially, the idea is that if we know $d'$ and the resulting linear system is given by a square matrix, then uniqueness of solution of degree $d'$ guarantees that this matrix is invertible, and the solution can then be expressed as an algebraic circuit given by Cramer's rule. However, the constraint matrix of the system might not be square in general, and in this case, we first make the system square (deterministically) before applying Cramer's rule. The details can be found in \cref{lem:approx-root-minpoly}.

One added subtlety here is that in general, we would not know the exact $y$-degree of $g$. In this case, we try all values from $1$ to $d-1$, and for each one formally construct an algebraic circuit with divisions and unbounded depth (as noted earlier, the circuit for $\Phi$ can already have unbounded depth)  and include all such circuits in our final list of solutions.

\paragraph*{Differences with \cite{KRS23}:} Many of the high-level ideas in the algorithm outlined above are similar to those in a prior work of Kumar, Ramanathan and Saptharishi \cite{KRS23}, where deterministic subexponential time algorithms were given for computing constant-degree factors of constant-depth circuits. The similarities include the motivation for the problems as well as the high-level approach of understanding the structure of the PIT instances appearing in the randomized algorithms for multivariate factorization for these special cases carefully and derandomizing those steps. However, there are key technical differences. The main technical result of  \cite{KRS23} is an upper bound on the complexity of the resultant of a constant-depth circuit and a constant-degree polynomial. However, we are unable to prove such an upper bound in our case. Instead, we define the pseudo-resultant and observe that its circuit complexity is comparable to that of the given input polynomial and the factors we care about. Moreover, it helps us completely avoid working with the resultant of $g$ and $h$ or the discriminant of $f$, and lets us find a good starting point for Newton iteration in deterministic subexponential time. There are also some technical differences in how we go from approximate power series roots to actual factors of $f$, but the main takeaway from this work is perhaps the notion of pseudo-resultant and its properties. 

\paragraph*{Inability to prune the list: } Unlike the results in \cite{KRS23}, where only true low degree factors of the input polynomial are output, we end up with a list of circuits containing the circuits for all irreducible factors with small constant-depth circuits, but perhaps much more. This difference stems from the fact that in the algorithm of \cite{KRS23}, at an intermediate step a list of constant-degree polynomials (essentially expressed as a sum of monomials) is constructed, where similar to this work, the list contains all the true constant-degree factors, as well as some spurious factors. However, since in \cite{KRS23} every polynomial in this intermediate list is a constant-degree polynomial in the monomial basis, the list is pruned to true factors by checking if a polynomial in the list divides the given input polynomial, and whether they are irreducible (which can be done deterministically since the degree of the polynomial is a constant). However, in this work, the list  consists of unbounded depth algebraic circuits with division gates. For such circuits, we do not even know how to test if the circuit is well-defined, i.e., that there are no denominators that are identically zero. Hence, it is unclear to us if the list can be pruned to the true irreducible factors in a non-trivial way deterministically. Addressing this issue seems like a very interesting open problem. 
  
\section{Notations and preliminaries}

We shall briefly describe our notation and some basic definitions. For a more detailed preliminaries section, please refer to \cref{sec:prelims} and \cref{sec:building-blocks}.
\begin{itemize}

\item Throughout this paper, we work over the field $\Q$ of rational numbers. For some of the statements that are used more generally, we use $\F$ to denote an underlying field.  

\item We use boldface lower case letters like $\vecx, \vecy, \veca$ to denote tuples, e.g. $\vecx = (x_1, x_2, \ldots, x_n)$. The arity of the tuple is either stated or will be clear from the context. 
\item For a polynomial $f$ and a non-negative integer $k$, $\homog_k(f)$ denotes the homogeneous component of $f$ of degree \emph{equal} to $k$. $\homog_{\leq k}(f)$ denotes the sum of homogeneous components of $f$ of degree at most $k$, i.e.,
\[
\homog_{\leq k}(f) := \sum_{i = 0}^k \homog_i(f)
\]

\item For a parameter $k\in \Z_{\geq 0}$, we will use $(\SP)^{(k)}$ to refer to product-depth $k$ circuits\footnote{We emphasize that this notation does \emph{not} refer to the $k^{\text{th}}$ power of a polynomial computed by a $\SP$ circuit.} with the root gate being $+$ and the deepest layer of gates being $\times$. Since any constant-depth algebraic circuit of depth $k$ and size $s$ can be converted to a formula of depth $k$ and size $s^{k+1}$ i.e. $\poly(s)$, we will use the terms circuits and formulas interchangeably, without any loss in the final bounds we prove.

\item Let $f$ and $g$ be multivariate polynomials such that $g \mid f$. Then, the \emph{multiplicity} or \emph{factor multiplicity} of $g$ in $f$ is defined to be the greatest integer $a$ such that $g^a$ divides $f$.

\item The size of a circuit over $\Q$ is equal to the sum of the number of edges and the sum of the bit-complexity of the constants appearing in the circuit. 

\item In our proofs, we encounter circuits over fields of the form $\Q(\alpha)$ where $\alpha$ is some algebraic number. For such circuits, we think of their description of each field constant as an element of $\mathbb{K} = \frac{\Q[u]}{A(u)}$ where $A(u)$ is its minimal polynomial of $\alpha$. The bit-complexity of any element of $\Q(\alpha)$ is defined as the bit-complexity of the unique representation as a element of $\mathbb{K}$. 
The bit-complexity of such a circuit is the total description size of the circuit, which includes the bit-complexity of any constants used internally.
\end{itemize}

\subsection{Newton iteration}

The following are some standard facts about Newton iteration and the proofs of these lemmas is present in \cref{app:newton-iteration-prelims}.

\begin{restatable}[Newton iteration]{lemma}{newtoniterationlemma}
  \label{lem:newton-iteration-approx-root}
  Let $F(\vecx,y) \in \F[\vecx,y]$ be monic in $y$. Suppose $u \in \F$ such that 
\begin{align*}
  F(\mathbf{0},u) & = 0,\\
  (\partial_y F)(\mathbf{0},u) & \neq 0.
\end{align*}
Then, for all $k \geq 0$, there is an \emph{approximate root} polynomial $\Phi_k \in \F[\vecx]$ that satisfies
\begin{align*}
  F(\vecx, \Phi_k) & = 0 \bmod{\inangle{\vecx}^{k+1}},\\
  \Phi_k(\mathbf{0}) & = u.
\end{align*}
Furthermore, if we are provided $F$ a circuit of size and bit-complexity $s$, and also the element $u \in \F$ with bit-complexity at most $B$, we can output an algebraic circuit for the approximate root $\Phi_k(\vecx)$ of size at most $\poly(s,k)$ and bit-complexity $\poly(s,B,k)$. 
\end{restatable}

\begin{restatable}[Computing minimal polynomials of approximate roots]{lemma}{computingminimalpolyslemma}
  \label{lem:approx-root-minpoly}
  Let $\Phi_k(\vecx) \in \F[\vecx]$ be provided by an algebraic circuit of size and bit-complexity at most $s$. Suppose there exists a polynomial $G(\vecx, y) = \sum_{i=0}^{d}{G_{i}(\vecx)y^i}$ that is monic in $y$ and irreducible satisfying
  \begin{align*}
    G(\vecx, \Phi_k(\vecx)) & = 0 \bmod{\inangle{\vecx}^{k+1}},\\
    \deg G & \leq D,\\
    \deg_y G & = d,\\
    k & \geq 2 D d.
  \end{align*}
  Then, given $d$ and the circuit for $\Phi_k$, for every $i\in[d]$, we can compute an algebraic circuit (with only $\vecx$ variables and no $y$), with divisions, of size and bit-complexity at most $\poly(s,d,D)$ for the polynomial $G_i(\vecx)$. In particular, we can compute a circuit of size and bit-complexity $\poly(s,d,D)$ for $G(\vecx,y)$.
\end{restatable}

\subsection{Pseudo-quotients for preserving indivisibility}

We define the pseudo-quotient of a pair of polynomials $f$ and $g$, which will act as a proxy for the quotient of $f$ and $g$.

\begin{definition}[Pseudo-quotients, \cite{Forbes15}]
  \label{defn:pseudo-quotient}
  Let $f,g \in \Q[\vecx]$ be non-zero polynomials with $g(\veczero) = \beta \neq 0$. The \emph{pseudo-quotient of $f$ and $g$} is defined as
  \[
    \homog_{\leq D - d}\inparen{\inparen{\frac{f(\vecx)}{\beta}} \cdot (1 + \tilde{g} + \tilde{g}^2 + \cdots + \tilde{g}^{D - d})}
  \]
  where $D = \deg(f)$, $d = \deg(g)$ and $\tilde{g} = 1 - \frac{g}{\beta}$.

  More generally, if $\vecalpha \in \Q^n$ is such that $g(\vecalpha) \neq 0$, the \emph{pseudo-quotient of $f$ and $g$ around $\vecalpha$} is defined as $\tilde{Q}(\vecx - \vecalpha, y)$ where $\tilde{Q}(\vecx, y)$ is the pseudo-quotient of $f(\vecx + \vecalpha)$ and $g(\vecx + \vecalpha)$.
\end{definition}

Forbes~\cite{Forbes15} showed that the pseudo-quotient allows one to reduce divisibility testing to a polynomial identity test. 

\begin{theorem}[Divisibility testing to PIT \cite{Forbes15}]\label{thm: forbes div testing}
  Let $f(\vecx)$ and $g(\vecx)$ be non-zero polynomials over a field $\F$ such that $g(\vec0) = \beta \neq 0$. Then, $g$ divides $f$ if and only if $f(\vecx)-Q_{f,g}(\vecx)\cdot g(\vecx)$ is identically zero, where $Q_{f,g}$ is the pseudo-quotient of $f$ and $g$. 
\end{theorem}

\section{Pseudo-resultant}\label{sec:pseudo-resultant}

In this section, we will define the \emph{pseudo-resultant} of a polynomial $f$ and a factor $g$. The pseudo-resultant serves as a proxy for the resultant as it is easy to argue about its circuit complexity, and it suffices for our applications. This definition and its applications to polynomial factorization that we discuss later in the paper are essentially the main technical contributions of this paper. 

For a polynomial $f(\vecx, y)$, let $S_{f}(\vecx,y)$ denote $(\partial_y f(\vecx,y))^2$ and $S_{f(\veca,y)}(y)$ denotes $(\partial_y f(\veca,y))^2$.

\begin{definition}[Pseudo-resultant] \label{def:pseudo-resultant}
  Given $f,g  \in \F[\vecx,y]$, let $Q(\vecx,y)$ be the pseudo-quotient of $S_{f}$ and $g$ interpreted as univariates in $y$ (with coefficients in $\F(\vecx)$). Then, the \emph{pseudo-resultant} of $f$ and $g$ with respect to $y$, denoted by $R_{f,g}$, is a polynomial in $\F[\vecx,y]$ defined as
  \[ R_{f,g} := S_{f} - Q\cdot g. \qedhere \] 
\end{definition}
The variable $y$ is special in the above definition, and will remain so throughout the paper.

\sloppy
We make the following simple observation that summarizes a natural property of the pseudo-quotient and the pseudo-resultant under substitutions. 
\begin{observation}[Pseudo-quotients and pseudo-resultants under substitutions]
  \label{obs: divisibility-under-substitutions} 
  Suppose $f(\vecx,y), g(\vecx,y) \in \F[\vecx, y]$ are monic in $y$, and suppose $\veca \in \F^{\abs{\vecx}}$ such that $g(\veca,0) \neq 0$. Let $Q(\vecx,y)$ be the pseudo-quotient of $S_{f}$ and $g$ interpreted as univariates in $y$ (with coefficients in $\F(\vecx)$). Let $Q'(y)$ be the pseudo-quotient of $S_{f(\veca,y)}$ and $g(\veca, y)$. Let $R_{f(\veca,y),g(\veca,y)}:= S_{f(\veca,y)} - Q'\cdot g(\veca,y)$ be the pseudo-resultant of $f(\veca,y)$ and $g(\veca,y)$. Then:
  \begin{enumerate}
    \item $Q(\veca, y) = Q'(y)$.
    \item $R_{f,g}(\veca,y) = R_{f(\veca,y),g(\veca,y)}(y)$.
  \end{enumerate}
\end{observation}
\begin{proof} 
  For the first part, let $g(\vecx, y) = g_0(\vecx) + g_1(\vecx) y + \cdots + y^d$. By the definition of the pseudo-quotient, we have 
 \begin{align*}
    Q(\vecx,y) & = \homog_{\leq D - d}\insquare{\frac{S_{f}(\vecx,y)}{g_0(\vecx)}\inparen{1 + \tilde{g}(\vecx,y) + \tilde{g}(\vecx,y)^2 + \dots + \tilde{g}(\vecx,y)^{D-d}}},\\
    Q'(y) & = \homog_{\leq D - d}\insquare{\frac{S_{f(a,y)}(y)}{g_0(\veca)}\inparen{1 + \hat{g}(y) + \hat{g}(y)^2 + \dots + \hat{g}(y)^{D-d}}}
 \end{align*}
 where $\tilde{g}(\vecx,y) = 1 - \frac{g(\vecx,y)}{g_0(\vecx)}$ and $\hat{g}(y) = 1 - \frac{g(\veca,y)}{g_0(\veca)}$. It is evident that $\hat{g}(y) = \tilde{g}(\veca,y)$. 
 Since the operations of taking partial derivatives with respect to $y$ (in the definitions of $S_f$ and $S_{f(\veca,y)}$), taking homogeneous components with respect to $y$ and evaluating the $\vecx$ variables commute with each other, it follows that $Q(\veca,y)=Q'(y)$ as claimed.

 For the same reason, it follows that $S_f(\vecx,y) = S_{f(\veca,y)}(y)$. Combining this observation with the first part, we get $S_f(\veca,y) - Q(\veca,y) g(\veca,y) = S_{f(\veca,y)}(y) - Q'(y)g(\veca,y)$ i.e. $R_{f,g}(\veca,y) = R_{f(\veca,y),g(\veca,y)}(y)$.   
\end{proof}

The following lemma tells us how to use the pseudo-resultant to find a suitable starting point $\veca$ for Newton iteration.

  \begin{lemma}[Properties of pseudo-resultant]\label{lem: pseudo-resultant properties} 
    Suppose $\F$ is a field, $f \in \F[\vecx,y]$ is monic in $y$, and $f=g\cdot h$, where $g$ is irreducible and $g \nmid h$. Let $S_f(\vecx,y) = (\partial_y f(\vecx,y))^2$, let $R_{f,g}$ denote the pseudo-resultant of $f$ and $g$ with respect to $y$, and let $R_{f(\veca,y),g(\veca,y)}(y)$ denote the pseudo-resultant of $f(\veca,y)$ and $g(\veca,y)$ with respect to $y$. Then, the following properties hold.
    \begin{enumerate}
      \item $g \nmid S_f$ in $\F[\vecx,y]$ and $\F(\vecx)[y]$. Equivalently, $R_{f,g}(\vecx,y) \not\equiv 0$ in $\F(\vecx)[y]$.
      \item For any $\veca \in \F^n$ such that $R_{f,g}(\veca,y) \not\equiv 0$ in $\F[y]$, $\veca$ satisfies the following property: there exists $u \in \overline{\F}$ such that $f(\veca,u) = 0$, $h(\veca,u) \neq 0$ and $\partial_y f(\veca,u) \neq 0$. In particular, $u$ is a multiplicity-1 root of $f(\veca,y)$.
    \end{enumerate}
  \end{lemma}
  \begin{proof} 
  Since the polynomial $f$ is monic in $y$, we have that  $g$, $h$ and $S_f$ are all monic in $y$ as well. Thus, by Gauss' Lemma (\cref{lem: gauss lemma}), we get that divisibility in $\F[\vecx,y]$ is the same as that in $\F(\vecx)[y]$, so we focus on divisibility in $\F[\vecx, y]$.  
 
 From $f = g\cdot h$ and the product rule for derivatives, we get 
 \[
 \partial_y f = h \cdot \partial_y g  + g \cdot \partial_y h \, .
 \]  
Now, since $g$ is irreducible, we have that $g$ divides  $(\partial_y f)^2$ if and only if $g$ divides $\partial_y f$. This, in turn, happens if and only if $g$ divides $h \cdot \partial_y g$. But from the hypothesis of the lemma, we know that $g$ is irreducible, and does not divide $h$. Moreover, since the $y$-degree of $\partial_y g$ is less than the $y$-degree of $g$, we have that $g$ does not divide $\partial_y g$ either. Thus, from the irreducibility of $g$, it follows that $g$ does not divide the product $h \cdot \partial_y g$, and hence, $g$ does not divide  $(\partial_y f)^2$ (in the ring $\F[\vecx, y]$). As mentioned earlier, it now follows from Gauss' lemma (\cref{lem: gauss lemma}) that   $g$ does not divide  $(\partial_y f)^2$ (in the ring $\F(\vecx)[y]$) i.e. $g \nmid S_f$. Combining the definition of $R_{f,g}$ and Forbes' reduction of divisibility testing to PIT (\cref{thm: forbes div testing}), it follows that $R_{f,g}(\vecx,y) \not\equiv 0$ in $\F(\vecx,y)$. This completes the proof of the first item. 

We now proceed with the proof of the second item. Let $\veca$ be a setting of the $\vecx$-variables such that $R_{f,g}(\veca,y) = R_{f(\veca,y),g(\veca,y)}(y) \not\equiv 0$ (where the equality follows from \cref{obs: divisibility-under-substitutions}) or equivalently, $g(\veca, y) \nmid S_{f}(\veca,y)$ (by \cref{thm: forbes div testing} and the fact that $S_f(\veca,y) = S_{f(\veca,y)}(y)$). Let $g(\veca, y)$ factor as  $g(\veca,y) = \prod_{i}{(y-\alpha_i)^{e_i}}$ over the algebraic closure $\overline{\F}$ of the field $\F$. Let $g_1(y) = \prod_{i: e_i = 1}{(y-\alpha_i)}$ be the product of linear factors of multiplicity 1, and $g_2(y) = \prod_{i: e_i \geq 2}{(y-\alpha_i)^{e_i}}$ be the product of rest of the factors. We first claim that for every root $\alpha_i$ of $g_2(y)$ with multiplicity $e_i \geq 2$, it holds that $(y-\alpha_i)^{e_i} \mid (\partial_y g(\veca,y))^2$. Indeed, denote by $q(y)$, the polynomial $g(\veca,y)/(y-\alpha_i)^{e_i}$. Then, by the  product rule, we have
       \[\partial_y g(\veca,y) = e_i(y-\alpha)^{e_i-1}q(y) + \partial_y q(y) \cdot (y-\alpha_i)^{e_i} \, ,\]
       
 Since $e_i \geq 2$, we have that $2(e_i - 1) \geq e_i $. Thus, $(y-\alpha_i)^{e_i} \mid (\partial_y g(\veca,y))^2$ and more generally, since $(y-\alpha_i)^{e_i}$ and $(y-\alpha_j)^{e_j}$ are relatively prime whenever $\alpha_i \neq \alpha_j$, we get that $g_2(y) \mid (\partial_y g(\veca,y))^2$. 
 
 Furthermore, setting $\vecx$ to $\veca$ in the expression for  $S_f(\vecx, y)$, we get that  \[
 S_f(\veca, y) =  (\partial_y g(\veca, y) \cdot h(\veca, y))^2 + (g(\veca, y) \cdot \partial_y h(\veca, y) )^2 + 2\cdot(\partial_y g(\veca, y)) \cdot g(\veca, y) \cdot h(\veca, y) \cdot (\partial_y h(\veca, y)) \, . \]
 We note here that all the partial derivatives in the above expression are taken with respect to the variable $y$, whereas all the substitutions are happening with respect to the $\vecx$-variables, and thus these two operations commute with each other. From the form of the expression above, the definition of $g_2$ and our earlier observation that $g_2(y) \mid (\partial_y g(\veca,y))^2$, it immediately follows that $g_2(y) \mid S_f(\veca, y)$.
 
 Since $g(\veca,y)$ does not divide $S_f(\veca,y)$ but $g_2(y)$ does, it must be the case that $g_1(y)$ does not divide $S_f(\veca,y)$. Thus, $g_1(y)$ has degree at least one and at least one of its roots in $\overline{\F}$ is \emph{not} a root of $S_f(\veca,y)$. We claim that this root satisfies the requirements of being the $u\in \overline{\F}$ from the claim. To this end, we note that
      \begin{itemize}
        \item $S_f(\veca,u) \neq 0$ or equivalently, $\partial_y f(\veca,u) \neq 0$
        \item $f(\veca,u) = g(\veca,u)\cdot h(\veca,u) = g_1(u)\cdot g_2(u) \cdot h(\veca,u) = 0$ since we chose $u$ so that $g_1(u) = 0$
        \item $\partial_y f(\veca,u) = \partial_g(\veca,u)\cdot h(\veca,u) + \partial_h (\veca,u)\cdot g(\veca,u) = \partial_y g(\veca,u)\cdot h(\veca,u)$ by the previous bullet. But $\partial_y f(\veca,u) \neq 0$ and hence $h(\veca,u) \neq 0$. 
      \end{itemize}
      This completes the proof of the second item.
  \end{proof}

\subsection{A starting point for Newton iteration}

The next lemma proves that in the setting of constant-depth circuits, the complexity of the pseudo-resultant of $f$ and $g$ is comparable to the complexity of $f$ and $g$. Hence, if we want to find a point $\veca$ that will help us start Newton iteration using the sufficient condition from \cref{lem: pseudo-resultant properties} (property 2), then it is enough to consider points in a hitting set for constant-depth circuits.

\begin{lemma}\label{lem:starting-point-for-newton-iteration}
  Suppose $f \in \F[\vecx,y]$ is a polynomial of degree $D$ with the following properties.
  \begin{itemize}
    \item $f$ is monic in $y$.
    \item $f=g\cdot h$, where $g = \sum_{i}{g_i(\vecx)y^i}$ is a monic irreducible polynomial of degree $d$ such that $g \nmid h$, and $g_0(\vecx) \neq 0$.
    \item $f$ can be computed by a circuit in $(\SP)^{\inparen{\Delta_1}}$ of size at most $s_1$ and $g$ can be computed by a size $s_2$ circuit in $(\SP)^{\inparen{\Delta_2}}$ for some constants $\Delta_1, \Delta_2 \in \N$.  
  \end{itemize}
  Let $R_{f,g}(\vecx,y) \in \F(\vecx)[y]$ be the pseudo-resultant of $f$ and $g$ with respect to $y$. For $\Delta = \max(\Delta_1,\Delta_2)$, let $\mathcal{H}$ be a hitting set for $(n+1)$-variate $(\SP)^{(\Delta + 2)}$-circuits of size at most $11s_1s_2D^5$ and degree at most $5D^2$. Then, there exists a point $\tilde{\veca}=(a_1, \dots,a_n, a_{n+1}) = (\veca,a_{n+1}) \in \mathcal{H}$ such that $R_{f,g}(\veca,y) \not\equiv 0$.
\end{lemma}

\begin{proof}
Let $Q(\vecx,y)$ be the pseudo-quotient of $S_f$ and $g$ as univariates in $y$. Let the numerator and the denominator of $R_{f,g}(\vecx,y) = S_f(\vecx,y) - Q(\vecx,y)\cdot g(\vecx,y)$ be $P_{\num}(\vecx,y)$ and $P_{\den}(\vecx)$ respectively. The proof proceeds by bounding the complexity of the numerator and the denominator, and observing that the given hitting set suffices to find a non-zero for both of them. 
   
   \paragraph{Bounding the complexity of the denominator $P_{\den}(\vecx)$ :}$S_f(\vecx,y)$ and $g(\vecx,y)$ do not contribute to the denominator. The only contributions to $P_{\den}(\vecx)$ (from $Q(\vecx,y)$) are $g_0(\vecx)$ by the $\frac{S_f(\vecx,y)}{g_0(\vecx)}$ term, and $g_0(\vecx)^{D-d}$ from the powers of $\tilde{g}$. Thus, $P_{\den}(\vecx) = g_0(\vecx)^{D - d + 1}$. Since $g(\vecx,y)$ has a $(\SP)^{(\Delta_2)}$-circuit of size $s_2$, so does $g_0(\vecx)$ (obtained by just setting $y$ to zero). Thus, $P_{\den}(\vecx)$ has a $(\SP)^{(\Delta_2+1)}$-circuit of size at most $s_2 + D \leq s_2D$. The degree of $P_{\den}$ is clearly upper bounded by $D^2$. 
    
    \paragraph{Bounding the complexity of the numerator $P_{\num}(\vecx,y)$ :}

    The numerator of $Q(\vecx,y)$, denoted by $Q_{\num}(\vecx,y)$, is 
    \[
      Q_{\num}(\vecx,y) =  \homog_{D-d}\insquare{S_f(\vecx,y)\inparen{\sum_{i=0}^{D-d}{g_0^{D-d-i}(g_0-g)^i}}} \, .
    \]
    From the definition of $S_f$ and standard computation of partial derivatives (see \cref{lem: partial derivatives with respect to one variable}), $S_f(\vecx,y)$ has a $(\SP)^{(\Delta_1)}$-circuit of size $(10s_1D^3)$. $\sum_{i=0}^{D-d}{g_0^{D-d-i}(g_0-g)^i}$ has a $(\SP)^{(\Delta_2+1)}$-circuit of size $s_2D^3$. Taking its product with $S_f(\vecx,y)$ and using interpolation to get the homogeneous components of degree at most $D-d$ (\cref{lem:computing-hom-components}) gives us a $(\SP)^{(\Delta+1)}$-circuit of size $(10s_1D^5)$ for $Q_{\num}$. Thus, $P_{\num}(\vecx,y) = P_{\den}(\vecx)S_f(\vecx,y) - Q_{\num}(\vecx,y)g(\vecx,y)$ has a $(\SP)^{(\Delta+1)}$-circuit of size $10s_1s_2D^5$. From the expression of the numerator, we have that a crude bound on its degree is at most  $2D + D^2 + D^2 \leq 4D^2$. 

    \paragraph{The hitting set $\mathcal{H}$ suffices :}
    
    Since $P_{\num}(\vecx,y)$ and $P_{\den}(\vecx)$ have relatively small $(\SP)^{(\Delta+1)}$-circuits, then  their product has a $(\SP)^{(\Delta+2)}$-circuit of size at most $10s_1s_2D^5 + s_2D \leq 11s_1s_2D^5$ and its degree is at most $D^2 + 4D^2 = 5D^2$. Thus, a hitting set $\mathcal{H}$ for this class will contain a point $\tilde{\veca} = (a_1, \dots, a_n, a_{n+1}) = (\veca,a_{n+1})$ such that $P_{\num}(\tilde{\veca})\cdot P_{\den}(\veca)$ is non-zero. In particular, $\frac{P_{\num}(\veca,y)}{P_{\den}(\veca)} = R_{f,g}(\veca,y) \not\equiv 0$.

\end{proof}

\section{Generating a list of candidate factors}\label{sec:generating-list}

{\small
\begin{algorithm}[H]
  \onehalfspacing
  \caption{Computing a list of candidate irreducible factors of multiplicity-1, degree $d$ and computable by a depth $\Delta'$ circuit of size $m$, when the input polynomial has depth $\Delta$}
  \label{algo:multiplicity-one-irreducible-factors}
  \SetKwInOut{Input}{Input}\SetKwInOut{Output}{Output}

  \Input{A size parameter $m$ in unary; degree parameter $d \in \N$; a $(\SP)^\Delta$-circuit of size $s$, bit-complexity $t$, total degree $D$, computing a polynomial $f(\vecx, y) \in \Q[x_1, \dots,x_n, y ]$ that is monic in $y$.}
  \Output{A list $L = \inbrace{C_1, \dots, C_r}$ of algebraic circuits (with division gates) such that for every  irreducible factor $g(\vecx, y)$ of $f(\vecx)$ s.t. $g^2 \nmid f$ and $g$ has a size $m$ circuit of depth $\Delta'$, $y$-degree equal to $d$, there exists $i \in [r]$ such that $C_i \equiv g(\vecx)$.}
  \BlankLine

  Set the list $L' = \emptyset$.

  Compute hitting-set $H_1$ for $(n+1)$-variate $\inparen{\SP}^{(\max(\Delta, \Delta')+2)}$-circuits of size $(11smD^5)$ and degree $(5D^2)$ using \cref{thm: PIT from LST}. \label{algo-one: PIT construction}
  
  Let $H_2$ be the projection of the points in $H_1$ on the first $n$ coordinates. 
 
{
    \For{$\veca = (a_1, \ldots, a_n)\in {H_2}$}{

      $F(\vecx,y) := {f}(x_1 + a_1, \dots, x_n + a_n, y)$ 
      
      Factorise the polynomial $F(\mathbf{0}, y) \in \Q[y]$ into irreducible factors as
      \[
        F(\mathbf{0}, y) = \sigma \cdot F_1(y)^{e_1} \cdots F_l(y)^{e_l}.
      \]
      where $0 \neq \sigma \in \Q$ and each $F_i(y)$ is monic. \label{alg:mult-one:factorize-F}
      
      \ForAll{$F_i \text{ s.t. } e_i = 1$}{

        Set $\mathbb{K}:= \frac{\Q[u]}{\inangle{F_i(u)}}$. 

        \If{$\partial_y F(\mathbf{0}, u) = 0$}{Skip to the next $F_i$.}

        Use \cref{lem:newton-iteration-approx-root} to obtain a circuit for $\Phi_k(\vecx)$ such that $F(\vecx, \Phi_k) = 0 \bmod{\inangle{\vecx}^{k+1}}$
        for $k = 2D^2$. 

        Use \cref{lem:approx-root-minpoly} to obtain a circuit (with divisions) $C$ for the minimal polynomial $G(\vecx, y)$ for $\Phi_k$ modulo the ideal $\inangle{\vecx}^{k+1}$. 
        
        Apply the transformation $x_i \rightarrow x_i - a_i$ on $C$ to get a circuit $C'$. 

        Update $L' := L' \cup \{C'\}$.
      }
    }
  }

  \Return{$L'$}
\end{algorithm}
}

\begin{remark*}[Obtaining circuits over the base field $\Q$]
  Although the above algorithm outputs circuits over algebraic extensions of $\Q$ of the form $\mathbb{K} = \frac{\Q[u]}{A(u)}$, they can be transformed syntactically to circuits over the base field $\Q$, with only a polynomially large blow-up in size, via standard techniques (\cref{lem:circuit-over-base-field}).
\end{remark*}

For simplicity, \cref{algo:multiplicity-one-irreducible-factors} deals with the special case when the factors we care about have multiplicity one and a fixed degree $d$. Later, we will describe the final algorithm that will iterate over the values of the degree and multiplicity parameters, and run \cref{algo:multiplicity-one-irreducible-factors} as a subroutine.

We now bound the running time of the above algorithm, and prove its correctness. In the following subsection, we use this algorithm to prove the main theorem \cref{thm:intro:main}. 
\subsection{Proof of correctness of \cref{algo:multiplicity-one-irreducible-factors}}

\begin{theorem}[Correctness of \cref{algo:multiplicity-one-irreducible-factors}]\label{thm:analysis-of-algorithm-for-multiplicity-one-factors}
  Let $\Q$ be the field of rational numbers and $\epsilon > 0$, $\Delta, \Delta' \in \N$ be arbitrary constants. Let  $f(\vecx,y) \in \Q[\vecx,y]$ be any polynomial that is monic in $y$, is computed by a $(\SP)^{(\Delta)}$-circuit of size $s$, bit-complexity $B$, degree $D$ and let $g$ be an irreducible factor of multiplicity one of $f$ with $y$-degree exactly $d$ such that $g$ is computable by a $(\SP)^{(\Delta')}$-circuit of size $m$. 

If \cref{algo:multiplicity-one-irreducible-factors} is invoked on input $m, d$, a circuit for $f$ with size $s$, bit-complexity $t$, $D$, then the output is a list $L$ of circuits of size $\poly(s,m,D)$ and bit-complexity $\poly(s,t,m,D)$ such that $L$ contains at least one circuit that computes $g$. 

Moreover, the algorithm terminates in time $(O(smD^5)^{O(\tilde{\Delta})}\cdot n)^{O_{\epsilon}((smD^7)^\epsilon)}\cdot \poly(s,D,B)$, where $\tilde{\Delta} = \max(\Delta,\Delta')+2$. 
\end{theorem}
\begin{proof}
We start with the proof of correctness. Suppose $f = g \cdot h$ where $g$ is one of the irreducible factors of interest. Let $R_{f,g}$ be the pseudo-resultant of $f$ and $g$ with respect to $y$. From \cref{lem:starting-point-for-newton-iteration}, we get that there exists a point $(a_1, a_2, \ldots, a_n, a_{n+1})$ in the hitting set $H_1$ defined in the algorithm such that $R_{f,g}(a_1, \dots, a_n, y) \not\equiv 0$.  In other words, there is an $\veca = (a_1, a_2, \ldots, a_n) \in H_2$ such that $R_{f,g}(\veca,y) \not\equiv 0$; let us fix such an $\veca$. 

From this and the second item of \cref{lem: pseudo-resultant properties}, we get that there is a $u \in \overline{\Q}$ such that $g(\veca,u) = 0$, $h(\veca, u) \neq 0$ and $\partial_y(f)(\veca,u) \neq 0$. If $G(\vecx, y) = g(\vecx + \veca, y)$ and $H(\vecx, y) = h(\vecx + \veca, y)$, this $u$ must be a root of one of the $F_i$'s obtained in \cref{alg:mult-one:factorize-F}. Hence, $F(\mathbf{0},u) = f(\veca, u) = 0$ and $\partial_y F(\mathbf{0},u) = \partial_y f(\veca, u) \neq 0$. We now satisfy the hypothesis for Newton Iteration (\cref{lem:newton-iteration-approx-root}) and we can obtain a circuit for the approximate root $\Phi_k$ for $F$ satisfying $\Phi_k(\mathbf{0}) = u$. Since $F(\vecx, \Phi_k) = G(\vecx, \Phi_k) \cdot H(\mathbf{}, \Phi_k) = 0 \bmod{\inangle{\vecx}^{k+1}}$ and $H(\mathbf{0}, \Phi_k(\mathbf{0})) = h(\veca, u)$ is a nonzero scalar in $\mathbb{K}$, we have that $G(\vecx, \Phi_k) = 0 \bmod{\inangle{\vecx}^{k+1}}$. Therefore, \cref{lem:approx-root-minpoly} will yield a circuit (with divisions) for $G$. Undoing the initial translate via $x_i \rightarrow x_i - a_i$ yields a circuit for $g(\vecx, y)$. 

From the description of the algorithm, we get that time complexity is at most 
\[
T_0 + |H_2| \cdot D \cdot (T_1 + T_2 + T_3)
\]
where, $T_0$ is the time taken to computing an appropriate hitting set in \cref{algo-one: PIT construction}, $T_1 \leq \poly(D,B)$ is the time taken to factorize the polynomial $F(\veczero,y)$ (\cref{thm: LLL univariate factorization}), $T_2 \leq \poly(s,D,B)$ is the time taken for computing the approximate root (\cref{lem:newton-iteration-approx-root}) and computing the minimal polynomial (\cref{lem:approx-root-minpoly}). The size of $H_2$ is at most the size of $H_1$, and \cref{thm: PIT from LST} tells us that $|H_1| \leq ((11smD^5)^{O(\tilde{\Delta})}\cdot n)^{O_{\epsilon}((44smD^7)^\epsilon)}$ (where $\tilde{\Delta}=\max(\Delta,\Delta')+2$); moreover, \cref{thm: PIT from LST} tells us that $T_0$, the time-complexity of computing $H_1$ at \cref{algo-one: PIT construction}, has the same expression as the size of $H_1$. Thus, the total running time is $((11smD^5)^{O(\tilde{\Delta})}\cdot n)^{O_{\epsilon}((55smD^7)^\epsilon)}\cdot \poly(s,D,B)$. The bit-complexity bound follows as the only nontrivial constant added to each circuit is the algebraic number $u$ whose minimal polynomial is one of the factors of $F(\veczero, y)$ and hence has small bit-complexity as well. 
\end{proof}

\subsection{Proof of \cref{thm:intro:main}}
We now describe our final algorithm, and its analysis to complete the proof of \cref{thm:intro:main}. 

{\small
\begin{algorithm}[H]
  \onehalfspacing
  \caption{Computing a list of candidate irreducible factors computable by a depth $\Delta'$ circuit of size $m$, when the input polynomial has depth $\Delta$}
   \label{algo:all-irreducible-factors}
  \SetKwInOut{Input}{Input}\SetKwInOut{Output}{Output}

  \Input{A size parameter $m$ in unary; a $(\SP)^\Delta$-circuit of size $s$, bit-complexity $t$, degree $D$, computing a polynomial $f(\vecx) \in \Q[x_1, \dots,x_n ]$.}
  \Output{A list $L = \inbrace{C_1, \dots, C_r}$ of algebraic circuits (with division gates) such that for every irreducible factor $g(\vecx)$ of $f(\vecx)$ s.t. $g$ has a size $m$ circuit of depth $\Delta'$, there exists $i \in [r]$ such that $C_i \equiv g(\vecx)$.}
  \BlankLine

  Set the output list $L' = \emptyset$.

  Compute hitting-set $H_1$ for $n$-variate $\inparen{\SP}^{(\Delta)}$-circuits of size $(sD^3)$ and degree $D$ using \cref{thm: PIT from LST}. \label{algo:all-irreducible-factors: PIT call H1}
 
  \For{$\vecalpha \in H_1$}{
    \label{candidate-sparse-fact-alg : outer-for-loop monicness}

    Define $\tilde{f}(\vecx, y) := f(\vecx + \vecalpha \cdot y ) = f(x_1 + \alpha_1 y , \ldots, x_n + \alpha_n y )$

	\For{$i = 0 \ldots D-1$}{\label{algo:all-irreducible-factors:iter-mult} 
	
	Compute a circuit $C_i$ for the polynomial $\tilde{f}_i(\vecx, y) := \frac{\partial^i \tilde{f}}{\partial y^i}$ 
	
	\For{$d = 1 \ldots D-1$}{ \label{algo:all-irreducible-factors:iter-degree}
	
	Compute a list $L_{i,d}$ of candidate degree $d$ irreducible factors of multiplicity one of $\tilde{f}_i(\vecx, y)$ using \cref{algo:multiplicity-one-irreducible-factors}	\label{algo:all-irreducible-factors:algo-1-call}
	
	Set $L' := L' \cup L_{i,d}$
	}		
	}
	}

  Let $L$ be the list of circuits  obtained by setting $y$ to $0$ in circuits in $L'$. 
		
  \Return{$L$}
\end{algorithm}
}

\begin{theorem}[Correctness of \cref{algo:all-irreducible-factors}]\label{thm:analysis-of-algorithm-for-all-factors}
  Let $\Q$ be the field of rational numbers and $\epsilon > 0$, $\Delta, \Delta' \in \N$ be arbitrary constants. Let  $f(\vecx) \in \Q[x_1, \dots, x_n]$ be a polynomial computed by a $(\SP)^{(\Delta)}$-circuit of size $s$, bit-complexity $t$, degree $D$ and let $g$ be an irreducible factor of $f$  such that $g$ is computable by a $(\SP)^{(\Delta')}$-circuit of size $m$. 
  
  If \cref{algo:all-irreducible-factors} is invoked on input $m$, a circuit for $f$ with size $s$, bit-complexity $t$, $D$, then the output is a list $L$ of circuits of size $\poly(s,m,D)$ and bit-complexity $\poly(s,t,m,D)$ such that $L$ contains at least one circuit that computes $g$. 
  Moreover, the algorithm terminates in time $\inparen{((O(smD^8))^{O(\tilde{\Delta})}\cdot n)}^{O_{\varepsilon}(smD^{10})^{\varepsilon}}\cdot \poly(s,D,T)$, where $\tilde{\Delta} = \max\{\Delta,\Delta'\} + 2$. 
\end{theorem}

\begin{proof}
Let $D$ be the total degree of $f$. If we view $f(\vecx + \vecalpha \cdot y)$ as a formal polynomial in $y$, we get that the coefficient of $y^D$ in it is equal to the evaluation of the degree $D$ homogeneous component of $f$ on input $(\vecalpha)$. Moreover, from \cref{lem:interpolation-univariate}, we get that this homogeneous component has a $(\SP)^{(\Delta)}$-circuit of size at most $sD^3$. Therefore, there exists an $\vecalpha $ in the hitting set $H_1$ for which the polynomial $\tilde{f}(\vecx,y) = f(\vecx + \vecalpha \cdot y)$ has $y$ degree equal to $D$, and hence (up to multiplication by a non-zero field constant) is monic in $y$. 

We will focus on an arbitrary fixed irreducible factor $g \mid f$ and prove that there exists a circuit in the final output list $L$ that computes $g$. Let $i^*$ be the multiplicity of $g$ in $f$. Let the degree of $g$ be $d$. Then, $\tilde{g}(\vecx,y) = g(\vecx+\vecalpha\cdot y)$ will be a degree-$d$ factor of multiplicity one of $\tilde{f}_{i^*-1}(\vecx,y) := \frac{\partial^i \tilde{f}}{\partial y^i}$. By \cref{lem: partial derivatives with respect to one variable}, $\tilde{f}_{i^*-1}(\vecx,y)$ has a $(\SP)^{(\Delta)}$-circuit of size at most $sD^3$. Moreover, the transformation $\vecx \mapsto \vecx + \vecalpha\cdot y$ maintains the irreducibility of $g$ by \cref{lem: irreducibility-under-shifts}. Thus, in the $(i^*-1)^{th}$ iteration of \cref{algo:all-irreducible-factors:iter-mult} and the $d^{th}$ iteration of \cref{algo:all-irreducible-factors:iter-degree}, \cref{algo:all-irreducible-factors:algo-1-call} will compute a list of candidate factors that will include $\tilde{g}$ as guaranteed by \cref{thm:analysis-of-algorithm-for-multiplicity-one-factors}. Setting $y=0$ (which can be done because \cref{lem:approx-root-minpoly} guarantees that only the $\vecx$ variables show up in the denominators) to go from the list $L'$ to $L$ will give us a list that contains $\tilde{g}(\vecx,0) = g$.

From the description of the algorithm, the running time of \cref{algo:all-irreducible-factors} is at most \[T_0 + |H_1|\cdot D^2 \cdot T_1 \cdot \poly(s,D,t)\]
where $T_0$ is the time taken to compute the hitting set $H_1$, $T_1$ is the time taken to invoke \cref{algo:multiplicity-one-irreducible-factors} in \cref{algo:all-irreducible-factors:algo-1-call}, $H_1$ is the hitting set computed in \cref{algo:all-irreducible-factors: PIT call H1} using \cref{thm: PIT from LST}, the $D^2$ factor is to account for the two loops in \cref{algo:all-irreducible-factors:iter-mult} and \cref{algo:all-irreducible-factors:iter-degree}, and the $\poly(s,D,t)$ factor is to account for the rest of the steps such as computing $\tilde{f}$ from $f$, computing circuits for derivatives, obtaining a list of circuits by setting $y=0$, etc.
$T_1$ is the running time of the first algorithm on $\tilde{f}_{i^*-1}$ which has a $(\SP)^{(\Delta)}$-circuit of size $s' = sD^3$. Thus, from \cref{thm:analysis-of-algorithm-for-multiplicity-one-factors}, $T_1$ is at most $((11smD^8)^{O(\tilde{\Delta})}\cdot n)^{O_{\epsilon}((55smD^{10})^\epsilon)}\cdot \poly(s,D,t)$. Applying \cref{thm: PIT from LST} for a $(\SP)^{(\Delta)}$-circuit of size $sD^3$ and degree $D$ tells us that $((sD^3)^{O(\Delta)}\cdot n)^{O_{\epsilon}((sD^4)^\epsilon)}$ will be a bound on both $T_0$ as well as the size of $H_1$. Thus, the final time complexity is $\inparen{(11smD^8)^{O(\tilde{\Delta})}\cdot n}^{O_{\varepsilon}(55smD^{10})^{\varepsilon}}\cdot \poly(s,D,T)$. The bit-complexity bound follows from \cref{thm:analysis-of-algorithm-for-multiplicity-one-factors}.
\end{proof}

\bibliographystyle{customurlbst/alphaurlpp}
{\let\thefootnote\relax
\footnotetext{\textcolor{\gitinfonotecolour}{\gitinfonote \easteregg}
}}
\bibliography{crossref,references}
\appendix
\section{Preliminaries}
\label{sec:prelims}

\subsection{Standard preliminaries using interpolation}

\begin{lemmawp}[Univariate interpolation (Lemma 5.3 \cite{S15})]\label{lem:interpolation-univariate}
  Let $f(x) = f_0 + f_1 x + \cdots + f_d x^d$ be a univariate polynomial of degree at most $d$. Then, for any $0 \leq r \leq d$ and there are\footnote{In fact, for any choice of distinct $\alpha_0, \ldots, \alpha_d$, there are appropriate $\beta_{r0}, \ldots, \beta_{rd}$ satisfying the equation. If the $\alpha_i$'s are chosen to have small bit-complexity, we can obtain a $\poly(d)$ bound on the bit-complexity of the associated $\beta_{ri}$'s.} field constants $\alpha_0, \ldots, \alpha_d$ and $\beta_{r0},\ldots, \beta_{rd}$ such that 
  \[
    f_r = \beta_{r0} f(\alpha_0) + \cdots + \beta_{rd} f(\alpha_d).
  \]
  Furthermore, the bit-complexity of all field constants is bounded by $\poly(d)$. 
\end{lemmawp}

\begin{lemma}[Computing homogeneous components (Lemma 5.4 \cite{S15})]
  \label{lem:computing-hom-components}
  Let $f \in \Q[\vecx]$ be an $n$-variate degree $d$ polynomial. Then, for an $0 \leq i \leq d$, there are field constants $\alpha_0,\ldots, \alpha_{d}$ and  $\beta_{i0}, \beta_{id}$ of bit-complexity $\poly(d)$ such that 
  \[
    \homog_i(f) = \beta_{i0} f(\alpha_0 \cdot \vecx) + \cdots + \beta_{id} f(\alpha_d \cdot \vecx).
  \]
  In particular, if $f$ is computable by $(\SP)^{(k)}$-formulas of size / bit-complexity at most $s$ then $\homog_i(f)$ is computable by $(\SP)^{(k)}$-formulas of size / bit-complexity at most $\poly(s,d)$.
\end{lemma}

\begin{lemma}[Computing partial derivatives in one variable]
  \label{lem: partial derivatives with respect to one variable}
  Let $f \in \Q[\vecx]$ be an $n$-variate degree $d$ polynomial. Then, for an $0 \leq r \leq d$, there are field elements $\alpha_i$'s and $\beta_{ij}$'s in $\Q$ of bit-complexity $\poly(d)$ such that 
  \[
    \frac{\partial^r f}{\partial x_1^r} = \sum_{i=0}^{d} x_1^i \cdot \inparen{\beta_{i0} f(\alpha_0, x_2,\ldots, x_n) + \cdots + \beta_{id} f(\alpha_d, x_2,\ldots, x_n)}
  \]
  In particular, if $f$ is computable by $(\SP)^{(k)}$-formulas of size / bit-complexity at most $s$ then $\frac{\partial^r f}{\partial x_1^r}$ is computable by $(\SP)^{(k)}$-formulas of size at most $8 s d^3$ and bit-complexity at most $\poly(s,d)$. 
\end{lemma}
\begin{proof}
  We may consider the polynomial $f$ as a univariate in $x_1$, and extract each coefficient of $x_1^i$ using \cref{lem:interpolation-univariate} and recombine them to get the appropriate partial derivative. That justifies the claimed expression. 

  As for the size, note that multiplying a $(\SP)^{(k)}$-formula of size  $s$ by $x_1^i$, by using distributivity of the top addition gate, results in a $(\SP)^{(k)}$-formula of size at most $s \cdot d$. Thus, the overall size of the above expression for the partial derivative is at most $8 s d^3$. 
\end{proof}

\subsection{Deterministic PIT for constant-depth circuits}

\begin{theorem}[PIT for constant-depth formulas (modification of Corollary 6 \cite{LST21})]\label{thm: PIT from LST}
  Let $\epsilon > 0$ be a real number and $\F$ be a field of characteristic 0. Let $C$ be an algebraic formula of size and bit-complexity $s \leq \operatorname{poly}(n)$, depth $k = o(\log\log\log n)$ computing a polynomial on $n$ variables, then there is a deterministic algorithm that can check whether the polynomial computed by $C$ is identically zero or not in time $(s^{O(k)}\cdot n)^{O_{\epsilon}((sD)^\epsilon)}$.
  \end{theorem}
  
\section{Building blocks}\label{sec:building-blocks}

\subsection{Univariate factorization over rational numbers}

The following classical theorem of Lenstra, Lenstra and \Lovasz{} gives us an efficient algorithm for factoring univariate polynomials over the field of rational numbers. 
\begin{theorem}[Factorizing polynomials with rational coefficients \cite{LLL82, GG13}]\label{thm: LLL univariate factorization}
  Let $f \in \Q[x]$ be a monic polynomial of degree $d$. Then there is a deterministic algorithm computing all the irreducible factors of $f$ that runs in time $\poly(d, t)$, where $t$ is the maximum bit-complexity of the coefficients of $f$. 
\end{theorem}

\subsection{Resultant}

\begin{definition}[The Resultant]\label{defn: resultant}
  Let $\mathcal{R}$ be a commutative ring. Given polynomials $g$ and $h$ in $\Ring[y]$, where:
  \begin{align*}
    g(y) & = g_0 + \cdots + y^d \cdot g_d \\
    h(y) & = h_0 + y \cdot h_1  + \cdots + y^D \cdot h_D
  \end{align*}
  with $g_d$ and $h_D \neq 0$ the \emph{Resultant} of $g$ and $h$, denoted by $\Res{y}{g}{h}$, is the determinant of the $(D + d) \times (D + d)$ Sylvester matrix $\Gamma$ of $g$ and $h$, given by: 
  \[\Gamma = 
    \begin{bmatrix}
        h_0  & h_1    & \dots  &        & h_{D} &        &       \\
            & \ddots & \ddots &        & \ddots & \ddots &       \\
            &        & h_0    & h_1    &        &  \dots & h_{D} \\
      g_{0}  & \dots  &        & g_{d}  &        &        &       \\
            &g_{0}   & \dots  &        & g_{d}  &        &       \\
            &        & \ddots & \ddots &        & \ddots &       \\
            &        &        & g_{0}  & \dots &        &g_{d} 
    \end{bmatrix}
  \]
      \par\vspace{-1.6\baselineskip}
\qedhere
\end{definition}

\begin{lemma}[Resultant and $\gcd$ (Corollary 6.20 \cite{GG13})]\label{lem: resultant gcd property}
  Let $\Ring$ be a unique factorization domain and $g, h \in \Ring[y]$ be non-zero polynomials. Then: $$\deg_y(\gcd(g,h)) \geq 1 \iff \Res{y}{g}{h} = 0$$
  where $\gcd(g,h) \in \Ring[y]$ and $\Res{y}{g}{h} \in \Ring$.
  
  Moreover, there exist polynomials $A, B \in \Ring[y]$ such that $\Res{y}{g}{h} = Ag + Bh$. 
\end{lemma}

\subsection{Irreducible polynomials in the field of fractions of a UFD}
\begin{lemma}[Gauss' Lemma (Corollary 6.10 \cite{GG13})]\label{lem: gauss lemma}
  Let $R$ be a unique factorization domain with field of fractions $K$. Suppose a polynomial $f\in R[y]$ is monic in $y$. Then $f$ is irreducible in $K[y]$ if and only if $f$ is irreducible in $R[y]$. 

  As a corollary, the factorization of any monic polynomial $f$ into its irreducible factors in $R[y]$ is exactly the factorization of $f$ into its irreducible factors in $K[y]$.
\end{lemma}

\subsection{Newton iteration}\label{app:newton-iteration-prelims}

\newtoniterationlemma*

\begin{proof}
  The approximate roots are given by the following recursive definition:
  \begin{align*}
    \Phi_0 & = u\\
    \text{For all $k \geq 0$,}\quad \Phi_{k+1} & = \Phi_k - \frac{F(\vecx, \Phi_k)}{\partial_y F (\mathbf{0}, u)}.
  \end{align*}
  It is clear from the above definition that $\Phi_{k+1}$ is a polynomial in $\vecx$ and its circuit complexity is at most an additive $O(s)$ larger than the circuit size of $\Phi_k$. Since the only constant introduced in the circuit is $\inparen{\partial_y F(\mathbf{0}, u)}^{-1}$, the bit-complexity is bounded by an additive $\poly(s,B)$. 

  We now show that $\Phi_{k+1}$ satisfies the requirement by induction (the base case of $\Phi_0$ is trivial). By Taylor expansion around $(\vecx, \Phi_k)$, if $z \in \inangle{\vecx}^{k+1}$, we have
  \begin{align*}
    F(\vecx, \Phi_k + z) & = F(\vecx, \Phi_k) + z \cdot \partial_y F(\vecx, \Phi_k) \\
     & \quad + \inparen{\frac{z^2}{2!}} \partial_{y^2} F(\vecx, \Phi_k) + \cdots \\
    & = F(x, \Phi_k) + z \cdot \partial_y F(\vecx, \Phi_k) \bmod{\inangle{\vecx}^{k+2}} & \text{(since $z^2 \in \inangle{\vecx}^{2(k+1)} \subseteq \inangle{\vecx}^{k+2}$)}.
  \end{align*}
  Furthermore, since $z \in \inangle{\vecx}^{k+1}$, we have that 
  \[
    z \cdot \partial_y F (\vecx, \Phi_k) = z \cdot \partial_y F (\mathbf{0}, \Phi_k(0)) \bmod{\inangle{\vecx}^{k+2}}
  \]
  since the other terms from the second multiplicand only contribute higher degree terms in $\vecx$. Thus, 
  \begin{align*}
    F(\vecx, \Phi_k + z) & = F(x, \Phi_k) + z \cdot \partial_y F(\mathbf{0}, u) \bmod{\inangle{\vecx}^{k+2}},\\
        & = 0 \bmod{\inangle{\vecx}^{k+2}} \quad\text{when $z = - \frac{F(x, \Phi_k)}{\partial_y F(\mathbf{0}, u)}$},\\
    \implies F(\vecx, \Phi_{k+1}) & = 0 \bmod{\inangle{\vecx}^{k+2}}. \qedhere
  \end{align*}
\end{proof}

\begin{lemma}[Mininum polynomials of approximate roots]\label{lem:approx-roots-and-minimal-poly}
  Let $\F$ be a field of characteristic zero. Let $\phi(\vecx) \in \F[\vecx]$ and $G(\vecx, y) \in \F[\vecx,y]$ be polynomials such that $G$ is irreducible of $y$-degree $d_y$, $\vecx$-degree $d$, is monic in $y$ and  satisfies 
  \[ G(\vecx, \phi) \equiv 0 \mod \langle \vecx \rangle^k \, , \] 
  for some natural number $k$ greater than $2d_yd$.  If $H(\vecx, y) \in \F[\vecx,y]$ is a non-zero polynomial of $y$-degree at most $d_y$, $\vecx$-degree $d$,  is monic in $y$ and satisfies 
  \[ H(\vecx, \phi) \equiv 0 \mod \langle \vecx \rangle^k \, , \] 
   then, $H$ must be equal to $G$.
  \end{lemma}
  \begin{proof}
  We consider both $G$ and $H$ as univariates in $y$ with coefficients in $\F(\vecx)$. Let $R$ be their resultant with respect to the variable $y$. Clearly, from the definition of the resultant, we have that $R$ is a polynomial in $\F[\vecx]$ of degree at most $2d_yd$. If $R$ is identically zero, then we have from \cref{lem: resultant gcd property} that $G$ and $H$ have a non-trivial GCD as polynomials in $\F(\vecx, y)$. But then, since $G$ and $H$ are monic in $y$, it follows from \cref{lem: gauss lemma} that they must have a non-trivial GCD in $\F[\vecx, y]$. Now, since $G$ is irreducible, this can happen only if $G$ divides $H$. Since the degree of $H$ is at most $d$, we get that they must be a non-zero constant multiple of each other. But they are both monic in $y$ and hence they must be equal to each other.  Thus, if $R$ is identically zero, then we are done. It now remains to show that $R$ is indeed identically zero. 
  
  We have from \cref{lem: resultant gcd property} that there exist polynomials $A(\vecx, y), B(\vecx, y)$ such that 
  \[
  A(\vecx, y)G(\vecx, y) + B(\vecx, y)H(\vecx, y) = R(\vecx) \, .
  \]
  Now, plugging in $y = \phi(\vecx)$ on both sides of the above equation, we get 
  \[
  A(\vecx, \phi(\vecx))G(\vecx, \phi(\vecx)) + B(\vecx, \phi(\vecx))H(\vecx, \phi(\vecx)) = R(\vecx) \, .
  \]
  But from the hypothesis of the lemma, we know that $G(\vecx, \phi) \equiv 0 \mod \langle \vecx \rangle^k$ and $H(\vecx, \phi) \equiv 0 \mod \langle \vecx \rangle^k$. Moreover, $k > 2d_yd$. Thus, the left-hand side of the identity above vanishes modulo $\langle \vecx \rangle^{2d_yd + 1}$. So, the resultant $R$ vanishes modulo $\langle \vecx \rangle^{2d_yd + 1}$. But since its degree is less than $2d_yd$, this means that $R$ is identically zero as a polynomial in $\F[\vecx]$.  This completes the proof of the lemma.   
  \end{proof}

\computingminimalpolyslemma*

\begin{proof}
  By \cref{lem:approx-roots-and-minimal-poly}, we know that $G$ must be the lowest degree polynomial of degree at most $D$ and $y$-degree at most $d$ that satisfies $G(\vecx, \Phi_k) = 0 \bmod{\inangle{\vecx}^{k+1}}$. We shall express this as a linear system in terms of the coefficients of $G$. 
  \begin{align*}
    G(\vecx, y) &= G_0 + G_1 \cdot y + G_2 \cdot y^2 + \cdots + G_{d} y^d\\
     &= \sum_{i=0}^d G_i y^i\\
     &= \sum_{i=0}^d \inparen{\sum_{j=0}^{D} G_{ij}} \cdot y^i
  \end{align*}
  where each $G_{ij}$ is the degree-$j$ homogeneous part of the coefficient of $y^i$ in $G$. We will treat each $G_{ij}$ as an indeterminate and express the condition $G(\vecx, \Phi_k(\vecx))$ as a system of linear equations in $G_{ij}$'s.\\

  For $r = 0,\ldots, k$, let $\Phi_k^r = \Phi_k^{(r,0)} + \cdots + \Phi_k^{(r,k)}$ where $\Phi_k^{(r,\ell)}$ is the degree-$\ell$ homogeneous part of $\Phi_k^r$. Given a circuit of size $s$ for $\Phi_k$, we can obtain $\poly(s)$-sized circuits for each of $\Phi_k^{(r,\ell)}$. 
  \begin{align*}
    G(\vecx, \Phi_k) & = \sum_{i=0}^d \sum_{j=0}^D G_{ij} \Phi_k^i = \sum_{i=0}^d \sum_{j=0}^D \sum_{\ell=0}^k G_{ij} \Phi_k^{(i,\ell)}\\
                     & = \sum_{m=0}^{D+k} \inparen{\sum_{i=0}^d \sum_{j=0}^D G_{ij} \Phi_k^{(i,m - j)}}
  \end{align*}
  where the last equality is just grouping the terms in terms of the $\vecx$-degree. Thus, the condition that $G(\vecx, \Phi_k(\vecx)) = 0 \bmod{\inangle{\vecx}^{k+1}}$ can be expressed as
  \begin{align*}
    \inparen{\sum_{i=0}^d \sum_{j=0}^D G_{ij} \Phi_k^{(i,m - j)}} & = 0 \quad \text{for all $m=0,\ldots, k$},\\
    G_{d,0} & = 1,\\
    G_{d,j} & = 0 \quad \text{for all $j > 0$}.
  \end{align*}
  From the uniqueness of solution, it follows that the column rank of the constraint matrix $M$ of this linear system is full. Now, if the coefficient vector of $G$ is a solution of $M \cdot \vecz = \vecb$, then it is also a solution of the linear system $M^{\dagger}\cdot M \vecz = M^{\dagger} \cdot \vecb$, where $M^{\dagger}$ denotes the conjugate transpose of $M$. Moreover, from \cref{lem:rank-of-M-transpose-M}, we have that $M^{\dagger}\cdot M$ is a square matrix of full rank. Thus, this unique solution of $M^{\dagger}\cdot M \vecz = M^{\dagger} \cdot b$ can be written in closed form as $(M^{\dagger}M)^{-1}(M^{\dagger}\cdot \vecb)$. 

  Now, by expressing the inverse $(M^{\dagger}M)^{-1} = \frac{\operatorname{adj}(M^\dagger M)}{\det(M^{\dagger} M)}$, where $\operatorname{adj}$ refers to the `adjugate matrix', we get a circuit of size (and bit-complexity) $\poly(s,d,D)$ for each $G_{ij}$, and hence, for $G(\vecx,y) = \sum_{i=0}^d \inparen{\sum_{j=0}^{D} G_{ij}} \cdot y^i$. 
\end{proof}

\subsection{A basic linear algebra fact}
We use the following basic fact in the proof of correctness of our algorithm. 
\begin{lemma}\label{lem:rank-of-M-transpose-M}
Let $M$ be an $r \times c$ matrix with entries in $\C[\vecx]$ such that $r \geq c$ and the rank of $M$ over the field $\C(\vecx)$ equals $c$. 

Then, the rank of the $c \times c$ matrix $M^{\dagger}M$ over the field $\C(\vecx)$ is also equal to $c$.  Here, $M^{\dagger}$ denotes the conjugate transpose of $M$. 
\end{lemma}
\begin{proof}

Since the rank of $M$ over $\C(\vecx)$ equals $c$, there is a substitution $\veca \in \C^n$ for the variables such that $N = M(\veca)$ has rank $c$ over $\C$. To show that $M^{\dagger}M$ is full rank over $\C(\vecx)$ it suffices to show that $N^{\dagger}N$ is full rank over $\C$. We do this by arguing that the kernel of $N^{\dagger}N$ does not contain a non-zero vector. 

Let $u \in \C^n$ be an arbitrary non-zero vector. We now argue that any such $u$ cannot be in the kernel of $N^{\dagger}N$. To see this, consider the inner product $\langle u, N^{\dagger}Nu\rangle$ where $\inangle{u,v} := u^{\dagger} v$. Clearly, this equals  $\langle Nu, Nu\rangle$, which equals the square of the $\ell_2$ norm of the vector $Nu$. Since $N$ has full column rank and $u$ us non-zero, we have that $Nu$ is a non-zero vector in $\C^n$, and thus its $\ell_2$ norm is non-zero. Thus, $N^{\dagger}Nu$ must be non-zero, and $N^{\dagger}N$ must be full rank.
\end{proof}

\subsection{Irreducibility under a shift of variables}

Since the final step of solving a linear system after Newton iteration works only when the candidate factor is irreducible, we need to ensure that the initial shift of variables does not affect the irreducibility of the factors; this is guaranteed by the following lemma.
\begin{lemma}\label{lem: irreducibility-under-shifts}
  Let $\F$ be a field and $g(\vecx) \in \F[\vecx]$ be an $n$-variate irreducible polynomial. Then, for every $\veca \in \F^n$, the polynomial $G(\vecx,y) := g(\vecx + \veca \cdot y)$ is also irreducible. 
  \end{lemma}

  \begin{proof}
    We will interpret the polynomial $g(\vecx) \in \F[\vecx]$ as $g(\vecx,y) \in \F[\vecx,y]$ with the property that it does not depend on the variable $y$. The mapping $(\vecx,y) \mapsto (\vecx + \alpha \cdot y,y)$ is an invertible linear transformation $M$ on the vector of variables $(x_1, \dots, x_n, y) = (\vecx,y)$. 
    
    We shall prove the contrapositive of the claim: if the polynomial $G(\vecx,y) := g(M(\vecx,y))$ has a non-trivial factorization, then so does $g$. Formally, if $g(M(\vecx,y)) = g_1(\vecx,y)\cdot g_2(\vecx,y)$ for some non-constant polynomials $g_1$ and $g_2$, then $g(\vecx,y) = g_1(M^{-1}(\vecx,y)) \cdot g_2(M^{-1}(\vecx,y))$ such that $g_1(M^{-1}(\vecx,y))$ and $g_2(M^{-1}(\vecx,y))$ are also non-constant polynomials. Observe that the operation of taking a product of two polynomials commutes with the operation of applying a linear transformation on the variables; it is easy to see for monomials, and it follows for polynomials by linearity of $M$. Thus, applying $M^{-1}$ on both sides of $g(M[\vecx,y]) = g_1(\vecx,y)\cdot g_2(\vecx,y)$ gives us that $g(\vecx,y) = g_1(M^{-1}(\vecx,y))\cdot g_2(M^{-1}(\vecx,y))$. Here, $g_1(M^{-1}(\vecx,y))$ and $g_2(M^{-1}(\vecx,y))$ will be non-constants since $M$ is invertible. Thus, if $G$ has a non-trivial factorization, then so does $g$.    
  \end{proof}
  
\subsection{Converting a circuit over a number field to a circuit over $\Q$}

\begin{lemma}[Constructing a circuit over the base field]
  \label{lem:circuit-over-base-field}
  Given as input an irreducible polynomial $A(u)$ of degree $r$ and bit-complexity $B$, and a $C$, with divisions, of size and formal-degree at most $s$ over a field $\frac{\Q[u]}{A(u)}$ that computes a polynomial $g(\vecx) \in \Q[\vecx]$. There is a deterministic algorithm that can output another circuit $C'$ with size $\poly(s,r,B)$ over the field $\Q$ computing the polynomial $g(\vecx)$. 
\end{lemma}
\begin{proof}
  \newcommand{\Cnum}{C_{\mathrm{num}}}
  \newcommand{\Cden}{C_{\mathrm{den}}}

  For any polynomial $p(u) \in \Q[u]$, we will use $(p \bmod A(u))$ to denote the unique polynomial $p'(u)$ of degree less than $A(u)$ such that $A(u) \mid p(u) - p'(u)$. 

  We may assume without loss of generality\footnote{The standard `division-elimination' approach keeps tracks of numerators and denominators in gate of the original circuit. This transformation only increases the size by a polynomial factor.} that the input circuit $C$ only has a division gate at the root, i.e. it computes $g$ in the form $g(\vecx) = \frac{\Cnum(\vecx)}{\Cden(\vecx)}$ where $\Cnum$ and $\Cden$ are circuits without divisions over $\Q[u]/A(u)$. Let $s, d$ be bounds on the size and degree of the circuits, respectively, of $\Cnum$ and $\Cden$. 

  By interpreting $u$ as a formal variable, let $\Cnum',\Cden' \in Q[\vecx]$ be the resulting polynomials (so that $\Cnum' \bmod A(u) = \Cnum$ and similarly for $\Cden'$). We may write $\Cnum', \Cden'$ as
  \begin{align*}
    \Cnum' & = \Cnum'^{(0)} + \Cnum'^{(1)} u + \cdots + \Cnum'^{(d)} u^d \in \Q[\vecx, u]\\
    \Cden' & = \Cden'^{(0)} + \Cden'^{(1)} u + \cdots + \Cden'^{(d)} u^d \in \Q[\vecx, u].\\
    \Cnum = (\Cnum' \bmod A(u)) & = \Cnum^{(0)} + \Cnum^{(1)} u + \cdots + \Cnum^{(r-1)} u^{r-1}\\
           & = \sum_{i=0}^d \Cnum'^{(0)} (u^i \bmod A(u))\\
    \Cden = (\Cden' \bmod A(u)) & = \Cden^{(0)} + \Cden^{(1)} u + \cdots + \Cden^{(r-1)} u^{r-1}\\
           & = \sum_{i=0}^d \Cden'^{(0)} (u^i \bmod A(u))
  \end{align*}
  Using \cref{lem:interpolation-univariate}, we can obtain circuits for each $\Cden'^{(i)}, \Cden'^{(i)}$ of size at most $\poly(s,d)$, and thus we can compute circuits for $\Cnum^{(i)}$ and $\Cden^{(i)}$ as well. 

  To `invert' $\Cden$, we consider indeterminates $b_0, b_1, \ldots, b_{r-1}$ such that 
  \[
    (\Cden^{(0)} + \cdots + \Cden^{(r-1)} u^{r-1}) \cdot (b_0 + b_1 u + \cdots + b_{r-1} u^{r-1}) = 1 \bmod{A(u)},
  \]
  which is once again a linear system over the indeterminate $b_j$'s, with coefficients being linear combinations of the circuits $\Cden^{(i)}$'s. We can therefore express the $b_j$'s as efficient rational functions using the circuits $\Cden^{(i)}$'s via Cramer's rule. 

  Finally, if $\frac{\Cnum}{\Cden}$ was indeed a polynomial $g(\vecx) \in \Q[\vecx]$, we must have that 
  \[
    ((\Cnum \cdot (b_0 + b_1 u + \cdots b_{r-1} u^{r-1})) \bmod A(u)) = g(\vecx).
  \]
  Thus, we can once again compute the LHS above and return the circuit for the coefficient of $u^0$ in the above expression. The size and bit-complexity bounds follow readily from the description.
\end{proof}

\end{document}